# CardiSort: a convolutional neural network for cross vendor automated sorting of cardiac MR images

Ruth P Lim, MBBS (Hons), DMedSc[1,2]; Stefan Kachel, BEng[1]; Adriana DM Villa, MD, PhD[3]; Leighton Kearney, MBBS, PhD[1,4]; Nuno Bettencourt, MD, PhD[5]; Alistair A Young, PhD[3]; Amedeo Chiribiri, MD, PhD[3]; Cian M Scannell, PhD[3]

1. Austin Health, Melbourne, Australia
2. The University of Melbourne, Departments of Radiology and Surgery (Austin), Melbourne, Australia
3. Kings College London, School of Biomedical Engineering and Imaging Sciences, London, United Kingdom
4. I-MED Radiology, Melbourne, Australia
5. University of Porto, Cardiovascular R & D Unit, Porto, Portugal

**Corresponding Author:**
Ruth P Lim
Austin Health/ The University of Melbourne
145 Studley Rd
Heidelberg
Victoria 3084
AUSTRALIA
Email: ruthplim74@gmail.com
Telephone: +61 3 9496 5431
Fax +61 3 9459 2817

**Abstract**

**Objectives**

To develop an image-based automatic deep learning method to classify cardiac MR images by sequence type and imaging plane for improved clinical post-processing efficiency.

**Methods**

Multi-vendor cardiac MRI studies were retrospectively collected from 4 centres and 3 vendors. A two-head convolutional neural network ('CardiSort') was trained to classify 35 sequences by imaging sequence (n=17) and plane (n=10). Single vendor training (SVT) on single centre images (n=234 patients) and multi-vendor training (MVT) with multicentre images (n = 479 patients, 3 centres) was performed. Model accuracy was compared to manual ground truth labels by an expert radiologist on a hold-out test set for both SVT and MVT. External validation of MVT (MVT$_{external}$) was performed on data from 3 previously unseen magnet systems from 2 vendors (n=80 patients).

**Results**

High sequence and plane accuracies were observed for SVT (85.2% and 93.2% respectively), and MVT (96.5% and 98.1% respectively) on the hold-out test set. MVT$_{external}$ yielded sequence accuracy of 92.7% and plane accuracy of 93.0%. There was high accuracy for common sequences and conventional cardiac planes. Poor accuracy was observed for underrepresented classes and sequences where there was greater variability in acquisition parameters across centres, such as perfusion imaging.

**Conclusions**

A deep learning network was developed on multivendor data to classify MRI studies into component sequences and planes, with external validation. With refinement, it has potential to improve workflow by enabling automated sequence selection, an important first step in completely automated post-processing pipelines.

**Introduction**

Cardiac magnetic resonance studies are commonly performed for comprehensive anatomic, functional and quantitative assessment and are relatively complex to perform and interpret. Recently updated Society for Cardiovascular Magnetic Resonance (SCMR) guidelines advocate standardized acquisition and postprocessing to ensure study quality and reproducibility[1,2]. Manual post-processing is time intensive and can be prone to human error. Therefore, there are active efforts to automate a range of quantitative tasks including ventricular segmentation, myocardial tissue characterization and perfusion assessment[3-6]. If validated and made available to the clinical community, automated post-processing pipelines could aid efficiency and consistency of measurements for diagnosis, prognosis and treatment monitoring.

Accurate identification of individual cardiac MR sequences is an important first step in channelling images to the appropriate post-processing tool. In the clinic, sequence labelling currently depends upon saved scan protocols and/ or real-time annotations by the scanning MR technologist, and are thus subject to large variations across centres, making standardization difficult. Consistent automated sorting would facilitate fully automated post-processing, with a proposed clinical workflow provided in Figure 1. As well as allowing for prospective automated post-processing, sequence identification can be used to automatically curate large datasets for training deep learning models. This curation has traditionally relied upon expert manual labour that is time-consuming and costly[7,8], and an automated means of data curation would facilitate use of larger datasets for more robust tool development and validation.

**Figure 1.**

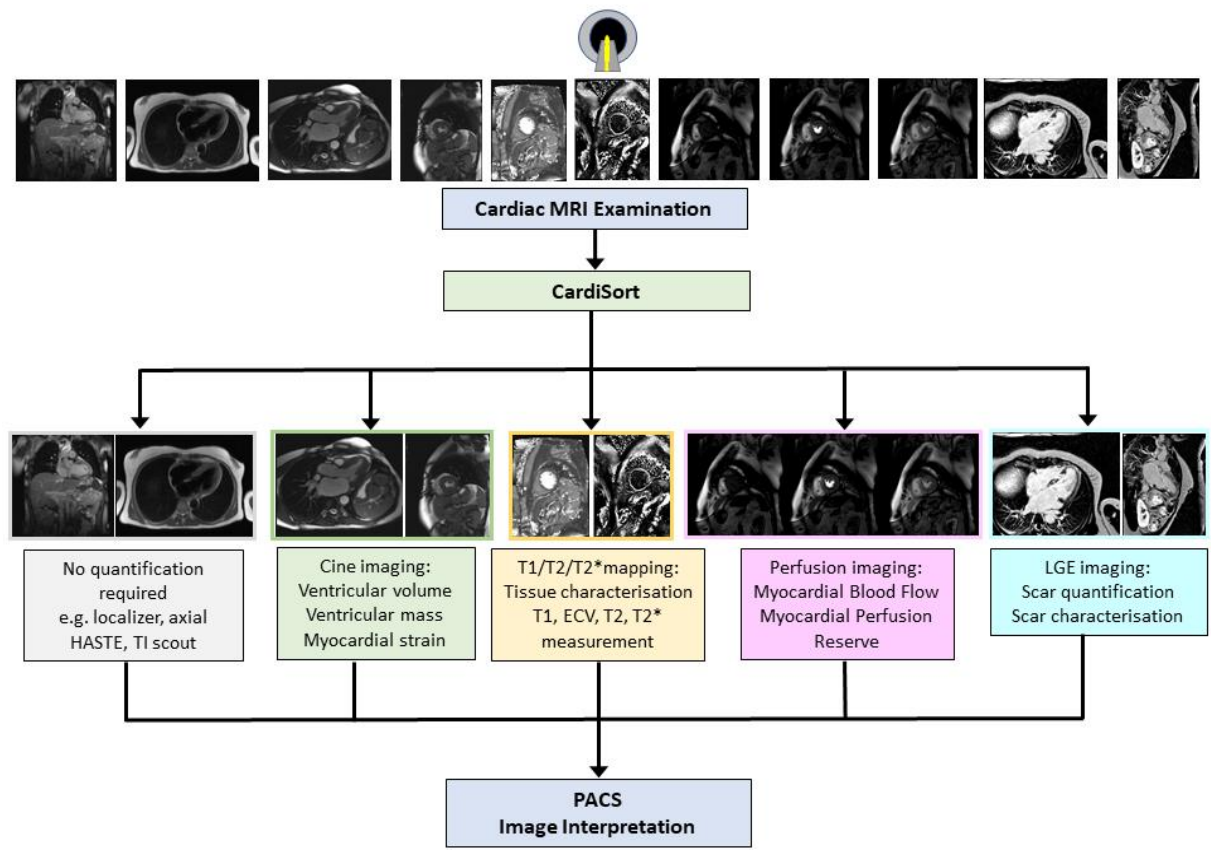

*Proposed streamlined clinical workflow for fully automated postprocessing, beginning with an automated cardiac sorting tool ('CardiSort'). CardiSort would receive images directly from the scanner and classify them, then automatically direct them for further quantitative post-processing as required. Those image types not requiring advanced post-processing would be sent directly to the Picture Archiving and Communication System (PACS). Some examples of multiple automated pipelines that could follow from the initial sorting step include: quantification of ventricular volume, function, mass and myocardial strain from cine imaging; extraction of T1, T2 and T2\* values and calculation of extracellular volume (ECV) from T1, T2 and T2\* sequences; measurement of myocardial blood flow and myocardial perfusion reserve from stress and rest perfusion imaging, and; quantification and characterisation of myocardial scar from late gadolinium enhanced (LGE) imaging. All images and extracted metrics would then be sent automatically to PACS for image interpretation by a cardiac MR radiologist or cardiologist.*

Van der Voort et al. described a convolutional neural network (CNN) approach for automated sorting of 8 brain MRI sequences, achieving greater than 98% accuracy in image labelling[9]. Cardiac MR image sorting is potentially more challenging, due to patient-specific cardiac planes, the plethora of sequences in the diagnostic armamentarium, and variability in sequence design and parameters, and it has not previously been studied in detail. The primary aim of this study was to develop a deep learning tool, CardiSort, to automatically classify a range of clinical cardiac MR sequences using pixel data alone, with real world cardiac MR data obtained across three different vendors. The model is also made available as an open-source tool for use by the community.

## Materials and Methods

### Study Design and Data

This was a retrospective study to create a model to classify 35 cardiac MRI sequences by sequence type and imaging plane. Anonymized cardiac MRI data was obtained from 4 centres and 3 vendors (Vendor 1, Philips; Vendor 2, Siemens; Vendor 3, General Electric) with institutional ethics approval. 334 randomly sampled patient studies (224M, 110F, mean±SD 54.8±15.8yrs) were obtained from Centre 1 from 2011 to 2020, with: a) 147 cases scanned at 1.5T (Ingenia, Philips Healthcare); b) 87 cases scanned at 3T (Achieva, Philips Healthcare), and; c) additional selected sequences obtained from 100 patients on a different 1.5T vendor system (Aera, Siemens Medical), to supplement training images for Vendor 2. Centre 1 common indications for scanning were ischaemic and non-ischaemic cardiomyopathy. 97 studies in patients with aortic stenosis (59M, 38F, 71.2±7.9y) were obtained from Centre 2, scanned at 1.5T from 2013 to 2017 (Symphony, Siemens Medical). 48 studies (33M, 15F, 60.0±17.3y) were obtained from Centre 3, scanned at 1.5T from 2017-2018 (Optima MR450w, GE Healthcare) with most common indications of non-ischaemic cardiomyopathy or arrhythmia.

Single vendor training (SVT) was firstly performed, utilizing only data from Centre 1, followed by multi-vendor training (MVT), utilizing data from Centres 1-3. The final SVT and MVT models were tested on hold-out test data from Centres 1-3.

For MVT external validation ($MVT_{external}$), the model was trained on data from Centres 1-3 as previously described. Testing was performed on 2 external datasets not previously seen by the model: Centre 2 data scanned in 2020 from a different vendor (Vendor 1) 3T system (Achieva, Philips, Healthcare), n=20, 14M, 6F, 58.9±13.5y; Centre 4 1.5T and 3T Vendor 2 data from 2016 to 2020 (Avanto and Skyra, Siemens Medical), n = 60, 31M, 29F, 55.2±14.6y. The most common clinical indications for external data were: non-ischaemic cardiomyopathy, assessment for arrhythmogenic foci, and myocarditis. Experiments performed are summarized in Figure 2.

*Pre-processing*

All images were obtained in Digital Imaging and Communications in Medicine (DICOM) format and anonymized. Secondary capture images were removed. Slightly different pre-processing using DICOM attributes[10] was required due to vendor-specific differences in the export of sequences with multiple image types, e.g., phase contrast imaging (PC), or multiple planes, e.g., cine imaging. For Vendors 1 and 3, all image types for such sequences were saved combined into a single series; series description, series instance unique identifier (series instance UID) and instance number were used for sorting. For Vendor 2, where multiple image types were saved in separate series but shared a protocol name, protocol name was also included for sorting. Data flow is summarized in Supplementary Figure 1, and sequence labels and final datapoints available are presented by individual magnet in Table 1.

**Figure 2.**

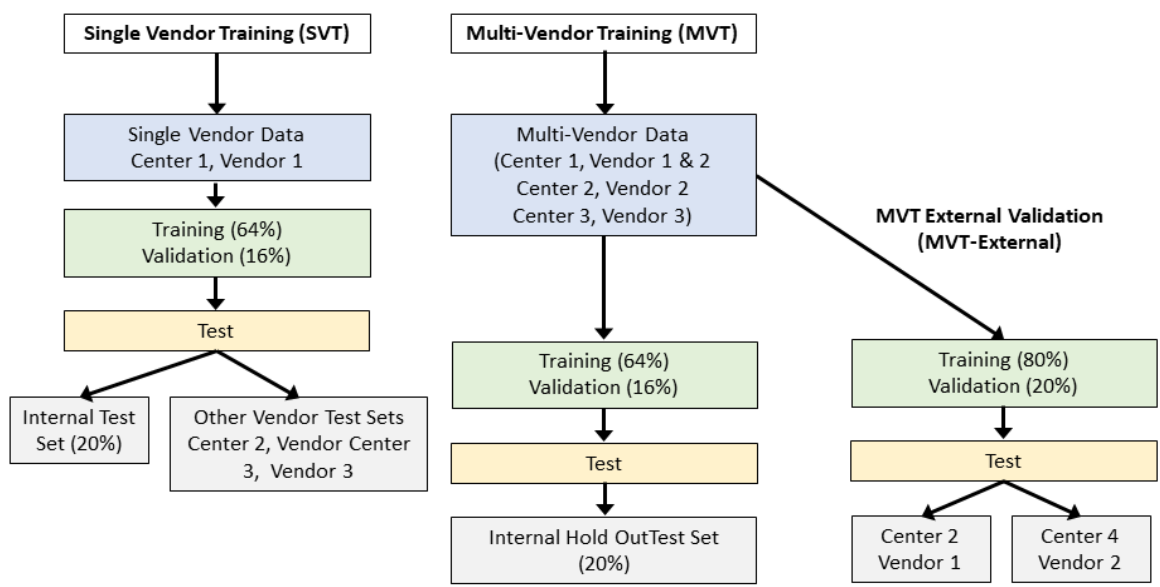

*Summary of experiments performed. Initially, a model was developed and trained on data from a single centre (Centre 1) and single vendor (Vendor 1), and tested on data from all 3 vendors. Subsequently, the model was trained on multi-vendor data from Centres 1-3 and tested on a hold-out test set derived from the same multi-vendor data. Finally, the multi-vendor trained model was tested on external data sets obtained from systems not used for training.*

For a unique image series, three images (first, middle, and last as sorted by position and instance number) were selected as input to the model. Three images were chosen as a fixed input size as required by the model. Although this discards images for some image series, it maintains a degree of the temporal or contrast information within the series. This was chosen empirically to balance model performance versus model size, based on preliminary experiments. Images were resized to a 256 x 256-pixel array, with array values normalized by the minimum and maximum values to between 0 and 1, per-channel. The three images were combined to form a "3-channel" array of shape 256 x 256 x 3 as a single datapoint for model input. For MRI sequences with less than 3 images, the first image was repeated once or twice as required.

**Ground Truth**

Ground truth labels were semi-automatically assigned. Series descriptions for each unique extracted image series were first assigned to classes by a board-certified cardiac radiologist with 15 years' cardiac MRI experience (SCMR Level 3 equivalent, anonymised for review). All data was then automatically sorted into separate classes based on Series Description. Classes were included for training if at least 20 unique datapoints for that class were present in the entire dataset, resulting in 35 labels incorporating both sequence type and imaging plane (Table 1). The remaining unassigned classes were excluded from further analysis.

Labelled images were then manually reviewed for completeness, label correctness and diagnostic quality by the same radiologist. Incorrectly labelled images were reassigned to the correct label if present, or excluded if absent. Aberrant congenital anatomy (n=1) and image sets considered non-diagnostic or incomplete were excluded from further analysis. The final datasets underwent repeat review for any misclassification by the same radiologist prior to network training, at least 4 weeks after initial review.

**Data Partitioning**

64% of experiment data was empirically chosen for training, 16% for validation and 20% for testing for SVT and MVT, with $MVT_{external}$ retrained on 80% and 20% of data for training and validation respectively. Stratified sampling of train, validation and test sets was performed to eliminate sampling bias, with study-level partitioning to ensure no closely related images from the training set were within validation or test sets.

**Model**

A 2D CNN, CardiSort, was iteratively developed to evaluate spatial imaging features for two-output multiclass classification. All inputs were shuffled prior to presentation to the network. An input layer and three deep convolutional layers with kernel sizes of 3 x 3 were employed for the model with 32, 32, 64 and 128 filters respectively. These were followed by two fully connected layers of 256 and 64 units respectively prior to the output layer, with separate outputs for sequence type and imaging plane (Figure 3). A ReLU activation function[11] was used for all layers prior to the output layer, with a softmax output used for classification[12]. He-normal weight initialization was employed [13].

**Figure 3.**

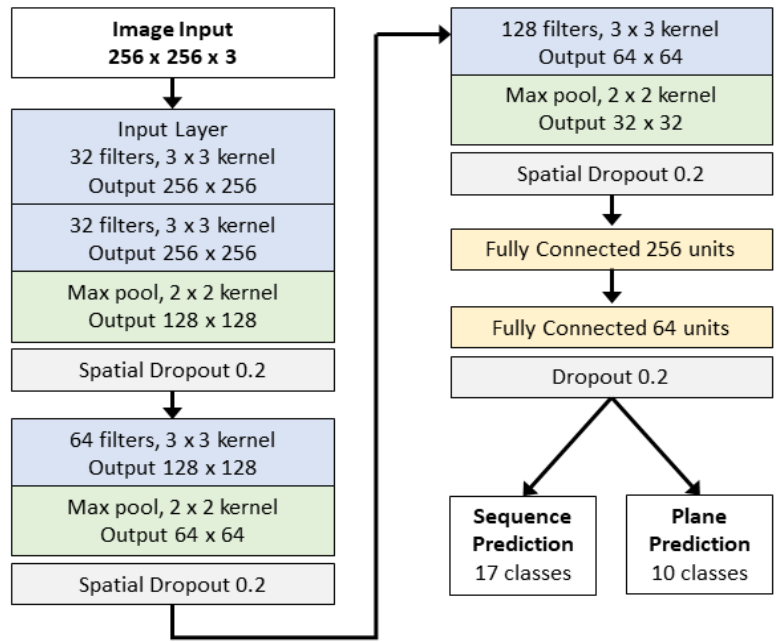

*Model Architecture. Batch normalization and ReLU activation was used for all hidden layers, with Softmax activation for classification.*

**Training**

*Data Augmentation*

For all experiments, minority classes were oversampled for the training dataset[14], such that the proportion of least to most common class representation was at least 1:4, empirically chosen to mitigate class imbalance whilst maintaining some real-world proportion of more versus less common sequences. For MVT and MVT$_{external}$, oversampling of training data was also performed to balance classes across vendors, with the proportion of least common to most common vendor empirically selected to be at least 1:2.

Random image augmentation was performed on the training dataset, employing alteration of image noise, image contrast, and composite affine and deformable geometric transformations[15]. Order of image channels for each training datapoint was also randomized.

*Hyperparameters*

The Adam optimizer was employed for training[16] ($\beta_1$ = 0.9 and $\beta_2$ = 0.999), with initial (minimum) and maximal learning rates determined with a learning rate finder[17], resulting in minimum learning rates of 2 x $10^{-4}$ (SVT, $MVT_{external}$) and 1 x $10^{-4}$ (MVT) and a maximum learning rate of 0.01 for all experiments. A cyclical triangular learning rate was used with a full learning rate cycle completed every 60 epochs [17].

Training was performed with a batch size of 32, and batch normalization[18]. Spatial dropout was employed after each max pooling layer for convolutional layers, and dropout after the two fully connected layers preceding the output layer, with a dropout rate of 0.2[19]. Training was performed for 480 epochs, with lowest summed validation loss observed at least 60 epochs prior to this for all experiments, using categorical cross entropy as the loss metric for both outputs.

The Keras environment with a TensorFlow backend (v2.5.0) was used for model development, with a GeForce RTX 3090 GPU (Nvidia Corporation, Santa Clara, USA) used for training. The MVT$_{external}$ model is made available at https://github.com/cianmscannell/cardisort, with accompanying code for its application to classifying non-curated data and sorting it into complete imaging series by sequence and plane label.

**Evaluation**

Model performance was assessed by overall and per-class classification accuracy for a) sequence type, b) imaging plane, and c) combined sequence and plane accuracy (Combined). Confusion matrices and Gradient-weighted class activation mapping (Grad-CAM) were employed to assess the developed models[20].

**Results**

For SVT, test set sequence type, plane and Combined accuracies were 85.2%, 93.2% and 81.8% respectively, with highest accuracy for Centre 1 data, and lower accuracy for Centres 2 and 3, where images were acquired on different vendor systems to the training data (Table 2, with per-class accuracies presented in Supplementary Table 1). An example of the differences (in white blood late gadolinium enhanced images (WBLGE)) between vendors, leading to poor accuracy on unseen vendors for SVT, is shown in Figure 4a.

MVT test set sequence type, plane and overall accuracies were 96.5%, 98.1% and 94.9% respectively (Table 2 and Supplementary Table 2). Excellent accuracy was observed for Vendor 2 data (98.7% sequence, 99.3% plane, 98.4% Combined), with near perfect sequence accuracy for Vendor 3 data (99.3% sequence, 96.2% plane, 95.5% Combined), and stronger plane versus sequence accuracy for Vendor 1 (95.2% sequence, 97.9% plane, 93.4% Combined). Lowest sequence accuracy was observed for Vendor 1 native modified Look Locker inversion recovery (MOLLI) imaging (39.0%), most frequently predicted as cine imaging, compared to 100% sequence accuracy for native MOLLI for Vendor 2, and no Vendor 3 MOLLI data available (Supplementary Table 2). Grad-CAM analysis demonstrated motion-related artefact as a cause of incorrect predictions for this sequence (Figure 4b). Fat-suppressed T2-weighted (FST2) weighted short axis imaging sequence accuracy was poor for Vendor 2 (4/6 datapoints, 66.7%), with failure of fat suppression and banding artefact observed (Figure 4c). MVT sequence and plane confusion matrices are presented in Figure 5a.

$MVT_{external}$ achieved sequence, plane and Combined accuracies of 92.7%, 93.0% and 86.6% on external data (Table 2, with per-class accuracies in Supplementary Table 3). Sequence, plane and Combined accuracies of 85.6%, 92.2% and 78.1% respectively for Vendor 1, and 96.1%, 93.4% and 90.8% respectively for Vendor 2 were recorded.

Excellent overall sequence accuracy was observed for cine (94-100%), phase contrast (100%) and long axis WBLGE (92-100%). Poorest sequence accuracy was observed for (Vendor 1) short axis perfusion imaging (0%), most commonly misclassified as cine (10/17 datapoints) or TI scout sequences (7/17). Differences in sequence parameters, total acquisition time and contrast protocol (dual-bolus[21] for Vendor 1 training versus single bolus technique for external data), leading to visually different image characteristics (Figure 4d). Relatively low sequence accuracy was also observed for WBLGE short axis imaging for Vendor 1 (16/21, 76.2%), classified as dark blood LGE imaging (DBLGE) in 4/21 cases. Suboptimal myocardial nulling and absence of myocardium and blood pool in the included magnitude image when positioned distal to the left ventricular apex was observed for incorrect predictions, impacting ability to differentiate between WBLGE and DBLGE (see Figure 4a, Vendor 1).

There was high plane prediction accuracy for most common sequence types including cine, MOLLI, TI scout and WBLGE imaging, strongest for 4 chamber (100%), axial (100%) and short axis imaging (304/323, 94.1%), and high for 2 chamber (115/129, 89.1%) and 3 chamber (100/113, 88.5%) planes. Poor plane performance was observed for the main pulmonary artery (1/10, 10%), 2 chamber FST2 (4/7, 57.1%), right ventricular outflow tract (5/10, 50%) and left ventricular outflow tract (9/13, 69.2%), with relatively few datapoints available for training. Poor plane performance for short axis FST2 (1/7, 14.3%) was noted, with differences in image export observed between external test (single slice location per series) and training data (multiple slice locations per series). Confusion matrices for sequence and plane are presented in Figure 5b.

**Figure 4.**

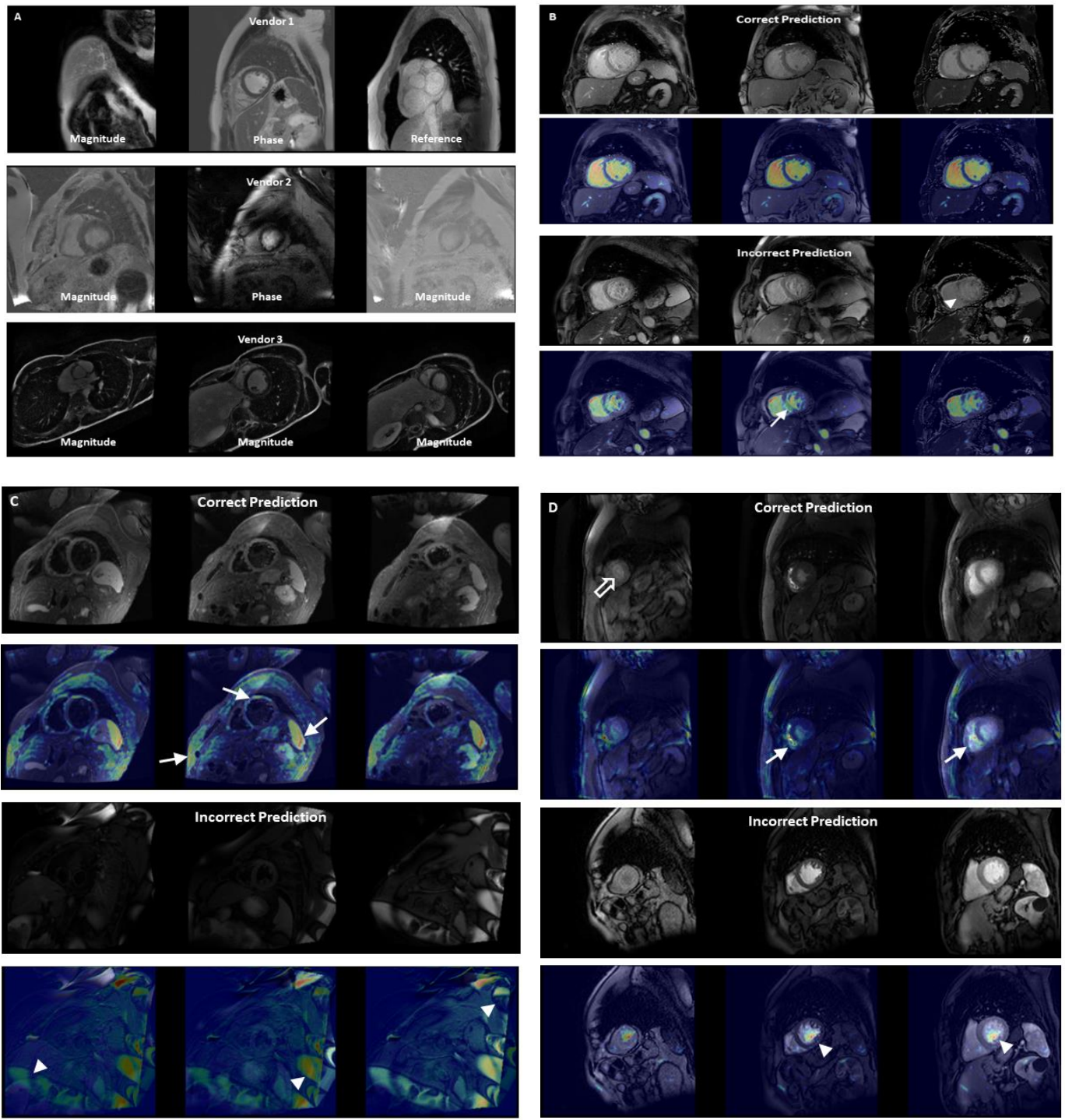

*Error Analysis. A) White blood LGE short axis imaging demonstrating differences between vendors, with 3 image types (magnitude inversion recovery (Magnitude), phase sensitive inversion recovery (Phase) and reference phase (Reference) images present for Vendor 1, Magnitude and Phase imaging for Vendor 2, and onlyMagnitude imaging present for Vendor 3 source data). Also note that differentiation between dark and white blood LGE imaging depends upon*

*appearance of blood and myocardium on magnitude images. B) Multi-vendor training native modified Look-Locker Inversion Recovery (MOLLI) T1 mapping, demonstrating a correct and incorrect prediction for Vendor 1 data. In both examples, model attention is focused upon the blood within the ventricles and blood vessels. However, motion-related artifact is present in the incorrect example, with activation visualized over the inferoseptal segment of the left ventricle (arrows), where myocardial signal appears similar to that of the blood pool on the T1 map (arrowhead), incorrectly predicted as cine short axis imaging. C) Multi-vendor training fat suppressed T2-weighted short axis imaging demonstrating a correct and incorrect prediction. For the correct prediction, model attention is focused upon the subcutaneous fat, myocardium and spleen (arrows), similar to structures a human reader would assess to identify the image type and plane. For the incorrect prediction, the model is focused upon banding artefact related to off-resonance effects at air/ soft tissue interfaces (arrowheads). This was predicted as cine short axis imaging, with cine imaging generally performed with balanced steady state free precession imaging, which is most prone to banding artefact. D) Multi-vendor training external short axis perfusion imaging examples of a correct prediction from the validation set and incorrect prediction from the external test set. For the correct prediction, model attention is focused upon the right ventricle (arrows) and to a lesser extent the left ventricle, where large relative fluctuations in the contrast of the blood pool of the left and right ventricular chambers are present. For the incorrect prediction, the model is more focused upon the blood pool of the left ventricle (arrowheads), with similar signal within left and right ventricles observed at the mid and basal levels. Note that apical blood pool signal was variable for both training/ validation and test dataset, with high signal sometimes observed secondary to flow-related "enhancement" on early perfusion imaging, prior to contrast arrival (hollow arrow).*

**Figure 5.**

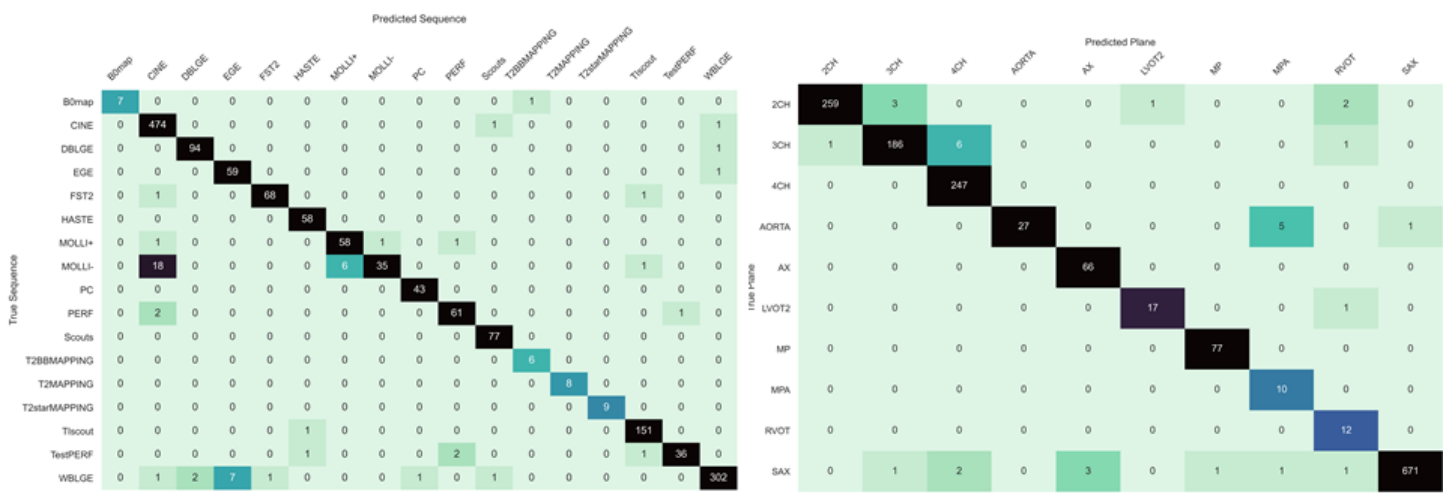

A. Hold-out Test Set

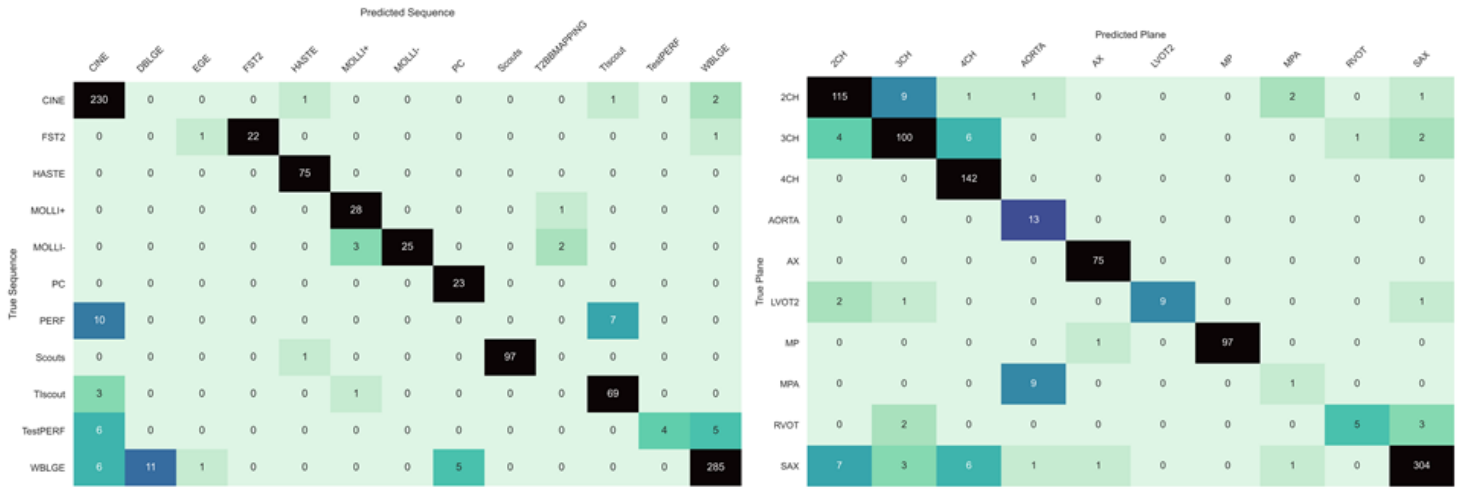

*Confusion Matrices for Multi-Vendor Training. A) Hold-out Test Set Sequence (left) and Plane (right) confusion matrices; B) External Data Sequence (left) and Plane (right) confusion matrices.*

**Discussion**

In this study, a multi-vendor deep learning model was developed and validated on external data with high accuracy for identifying commonly performed sequences, particularly cine and WBLGE imaging, and for standard anatomic and cardiac planes. Our work highlights the importance of training the model on a diverse, multi-institutional multi-vendor dataset, with SVT demonstrating high performance on hold-out test data from the same institution, vendor and magnet systems, but poor performance on data from other institutions and vendors, with improvement in accuracy with MVT on the hold-out multivendor test set.

The variability of clinically used cardiac MRI sequences is also highlighted, impacting model generalizability, exemplified in our study by the decrease in accuracy of MVT from internal test to external data. Model performance was lower for sequences with parameter differences between external validation and training images than more uniform sequences, shown by our experience with perfusion imaging. Image artefacts, e.g., misregistration from motion for MOLLI, banding artefact, or failure of fat suppression, also impacted model accuracy. Differences in image export between vendors and institutions also impacted the robustness of the developed model. Vendors provide users with the option to save and export a sequence with multiple slice locations as a single or multiple series. This impacted the model's ability to predict short axis FST2 for $MVT_{external}$, where differences in image storage existed between training and external data. Even for sequences that appear relatively uniform, subtle cross-vendor and within-vendor differences may impact model generalizability, as was observed for short axis cine imaging for automated ventricular segmentation[22]. Data augmentation can aid generalizability but does not entirely solve the problem. Greater exposure to a larger variety of data and permutations of image storage for classes with multiple slice locations would likely improve accuracy.

We have made CardiSort available for the scientific community to utilise, facilitating development of fully automated pipelines by allowing automated selection of the requisite image series for processing. We have also provided our trained model weights, should other scientists wish to use transfer learning to include their own data, additional classes, or other magnet systems. Manual selection of the desired sequence is still required in automated cardiac MRI post-processing pipelines described to date[4-6,23]. Automated extraction from non-curated data would provide a crucial first step to applying AI prospectively in the clinic, rather than retrospectively in a controlled environment. A proposed future clinical workflow would incorporate CardiSort to classify images received directly from the scanner, with sorted images then automatically sent to the appropriate post-processing pipeline for quantitative metrics, e.g., cine imaging for ventricular segmentation[24],myocardial strain[25], perfusion quantification [4], and scar quantification and classification[26], and results of post-processing automatically returned for image interpretation.

There are some limitations to our work. This was a relatively small dataset for a deep learning task, with data augmentation used to expose the network to a greater variety of inputs. Some classes were not available for every vendor in the training data and for external validation. We deliberately chose not to incorporate metadata within model input, which is variable between vendors/ centres and may be incomplete or unavailable in publicly available datasets. While the datasets reflect real world composition of clinical adult cardiac MRI, the model was only trained to recognize standard cardiac anatomy, with insufficient data available for congenital abnormalities.

Our approach, sampling only 3 images per sequence, likely impacted accuracy for predicting sequences that vary temporally, in image contrast or included image type throughout the acquisition, as was observed for short axis WBLGE for $MVT_{external}$. This represents a practical and efficient means of representing sequences to a model requiring data of a consistent shape for training. Cardiac MRI sequences may comprise a single image or multiple images, and a single or up to three image types, and this approach ensured representation of these. Future work might incorporate other architectures, for example, a 3-dimensional CNN with greater number of input images per sequence and/ or a recurrent neural network to incorporate more sequential information, at the expense of training efficiency.

In conclusion, we have trained a deep learning network on multi-institutional multi-vendor data to infer 35 unique cardiac MRI sequences by sequence type and imaging plane, with high performance for the most common image types. We have also made our work available for other scientists to use on their own non-curated datasets or to adapt to include additional sequences. Sorting of non-curated data represents a heretofore missing link in the development of efficient and fully automated processing pipelines, essential if they are to be ultimately translated from the research to the clinical domain.

**Funding**

This work was supported by the Wellcome/ EPSRC Centre for Medical Engineering [WT 203148/Z/16/Z]

**Tables**

**Table 1.** Sequence and plane of all data, with prevalence of datapoints presented by magnet.

|  |  | Data used for model training | | | | | External Validation | | |
|---|---|---|---|---|---|---|---|---|---|
|  |  | Centre 1 | | | Centre 2 | Centre 3 | Centre 2 | Centre 4 | Centre 4 |
| Vendor |  | Philips | Philips | Siemens | Siemens | GE | Philips | Siemens | Siemens |
| Field Strength (T) |  | 1.5 | 3 | 1.5 | 1.5 | 1.5 | 1.5 | 1.5 | 3 |
| Sequence | Plane |  |  |  |  |  |  |  |  |
| B0 map | Axial | 0 | 36 | 0 | 0 | 0 | 0 | 0 | 0 |
| Cine bSSFP | 2-chamber | 275 | 131 | 0 | 181 | 50 | 24 | 13 | 4 |
| Cine bSSFP | 3-chamber | 153 | 54 | 0 | 94 | 47 | 27 | 14 | 6 |
| Cine bSSFP | 4-chamber | 192 | 145 | 0 | 109 | 50 | 27 | 14 | 15 |
| Cine bSSFP | LVOT | 43 | 3 | 2 | 40 | 0 | 0 | 8 | 5 |
| Cine bSSFP | RVOT | 31 | 4 | 0 | 0 | 7 | 0 | 0 | 10 |
| Cine bSSFP | Short axis | 345 | 194 | 0 | 107 | 48 | 35 | 19 | 13 |
| DBLGE | 2-chamber | 127 | 3 | 0 | 0 | 0 | 0 | 0 | 0 |

| | | | | | | | | |
|---|---|---|---|---|---|---|---|---|
| **DBLGE** | **3-chamber** | 107 | 5 | 0 | 0 | 0 | 0 | 0 | 0 |
| **DBLGE** | **4-chamber** | 110 | 4 | 0 | 0 | 0 | 0 | 0 | 0 |
| **DBLGE** | **Short axis** | 129 | 3 | 0 | 0 | 0 | 0 | 0 | 0 |
| **EGE** | **2-chamber** | 112 | 3 | 0 | 0 | 0 | 0 | 0 | 0 |
| **EGE** | **3-chamber** | 83 | 3 | 0 | 0 | 0 | 0 | 0 | 0 |
| **EGE** | **4-chamber** | 84 | 3 | 0 | 0 | 0 | 0 | 0 | 0 |
| **FST2** | **2-chamber** | 26 | 3 | 25 | 0 | 0 | 3 | 3 | 1 |
| **FST2** | **3-chamber** | 22 | 0 | 27 | 0 | 0 | 0 | 3 | 1 |
| **FST2** | **4-chamber** | 28 | 3 | 25 | 0 | 47 | 2 | 3 | 1 |
| **FST2** | **Short axis** | 30 | 9 | 27 | 0 | 45 | 4 | 0 | 3 |
| **HASTE** | **Axial** | 137 | 11 | 0 | 88 | 47 | 16 | 45 | 14 |
| **MOLLI-** | **Short axis** | 136 | 66 | 94 | 0 | 0 | 0 | 30 | 0 |
| **MOLLI+** | **Short axis** | 141 | 53 | 113 | 0 | 0 | 0 | 29 | 0 |
| **Phase contrast** | **Aorta** | 31 | 7 | 0 | 103 | 11 | 0 | 1 | 12 |
| **Phase contrast** | **MPA** | 23 | 5 | 0 | 0 | 8 | 0 | 1 | 9 |
| **Perfusion** | **Short axis** | 60 | 83 | 0 | 175 | 0 | 17 | 0 | 0 |
| **Scout Imaging** | **Multiplanar** | 156 | 121 | 0 | 87 | 0 | 8 | 90 | 0 |
| T2 mapping bright blood | Short axis | 0 | 0 | 28 | 0 | 0 | 0 | 0 | 0 |
| T2 mapping dark blood | Short axis | 26 | 16 | 0 | 0 | 0 | 0 | 0 | 0 |
| T2* mapping | Short axis | 9 | 33 | 5 | 0 | 0 | 0 | 0 | 0 |
| Test Perfusion (Pre contrast) | Short axis | 36 | 55 | 0 | 93 | 0 | 15 | 0 | 0 |
| TI scout | 4-chamber | 30 | 0 | 5 | 0 | 0 | 0 | 0 | 0 |
| TI scout | Short axis | 271 | 55 | 172 | 90 | 52 | 18 | 43 | 12 |
| WBLGE | 2-chamber | 163 | 60 | 0 | 101 | 47 | 34 | 42 | 5 |
| WBLGE | 3-chamber | 141 | 55 | 0 | 98 | 47 | 25 | 33 | 4 |
| WBLGE | 4-chamber | 147 | 48 | 0 | 96 | 49 | 30 | 46 | 4 |
| WBLGE | Short axis | 221 | 78 | 0 | 176 | 47 | 21 | 50 | 14 |
| Total | | 3625 | 1352 | 523 | 1638 | 604 | 306 | 487 | 133 |

bSSFP = balanced steady state free precession imaging

LVOT = left ventricular outflow tract, perpendicular to 3-chamber plane

RVOT = right ventricular outflow tract (oblique sagittal plane)

DBLGE = dark blood late gadolinium enhanced images (blood pool nulled)

EGE = Early gadolinium enhanced images

FST2 = fat suppressed T2 weighted imaging

HASTE = Half Fourier Acquisition single shot turbo spin echo imaging

MOLLI+ = Modified Look Locker Inversion Recovery imaging post contrast

MOLLI- = native Modified Look Locker Inversion Recovery imaging

MPA = main pulmonary artery

TI scout = inversion time scout imaging for late gadolinium enhanced imaging

WBLGE = white blood late gadolinium enhanced images (normal myocardium nulled)

**Table 2.** Accuracy of model on test data for each experiment by Centre and Vendor. Overall accuracy represents data where combined sequence type and plane accuracy were correct.

| Experiment | Centre | Vendor | Sequence Accuracy (%) | Plane Accuracy (%) | Overall Accuracy (%) |
|---|---|---|---|---|---|
| **Single Vendor Training*** | 1 | Philips | 958/1025 (93.46) | 1004/1025 (97.95) | 938/1025 (91.51) |
| | 1 & 2 | Siemens | 322/438 (73.52) | 372/438 (84.93) | 295/438 (67.35) |
| | 3 | GE | 79/133 (59.40) | 112/133 (84.21) | 73/133 (54.89) |
| | All | All | **1359/1596 (85.15)** | **1488/1596 (93.23)** | **1306/1596 (81.83%)** |
| **Multi-Vendor Training** | 1 | Philips | 976/1025 (95.22) | 1003/1025 (97.85) | 957/1025 (93.37) |
| | 1 & 2 | Siemens | 438/444 (98.65) | 441/444 (99.32) | 437/444 (98.42) |
| | 3 | GE | 132/133 (99.25) | 128/133 (96.24) | 127/133 (95.49) |
| | All | All | **1546/1602 (96.50)** | **1572/1602 (98.13)** | **1520/1602 (94.88)** |
| **Multi-Vendor Training – External Validation** | 2 | Philips | 262/306 (85.62) | 282/306 (92.16) | 239/306 (78.10) |
| | 4 | Siemens | 596/620 (96.13) | 579/620 (93.39) | 563/620 (90.81) |
| | All external data | | **858/926 (92.66)** | **861/926 (92.98)** | **802/926 (86.61)** |

* One class (short axis T2 bright blood mapping) was omitted from test datasets for single vendor training due to absence of this class in the training data